\documentclass[aps,pra,reprint,superscriptaddress,floatfix]{revtex4-2}

\usepackage{amsmath,amssymb,bm}
\usepackage{braket}
\usepackage{graphicx}
\usepackage{microtype}

\newcommand{\ii}{\mathrm{i}}

\newcommand{\avg}[1]{\left\langle #1\right\rangle}
\newcommand{\abs}[1]{\left|#1\right|}

\begin{document}

\title{Dicke model of open-shell atoms with tensor polarizability}

\author{Ziheng Zhu}
\affiliation{Graduate School of China Academy of Engineering Physics, Beijing 100193, China}
\author{Yu Chen}
\affiliation{Graduate School of China Academy of Engineering Physics, Beijing 100193, China}

\begin{abstract}
We study the extended Dicke model obtained from the scalar--vector--tensor
polarizability of open-shell atoms coupled to a shared cavity mode.
For a single electric-dipole channel, rotational symmetry fixes the tensor to
vector ratio; we use $^{168}\mathrm{Er}$ with $J=6$ as a product-state
example.
A circular pump produces a chiral Raman vertex whose tensor-dependent matrix
elements select the occupied Zeeman sector.
The resulting first-order transition occurs before the empty-cavity branch
loses local stability, and the ordered state is read out through
$\langle\hat J_z\rangle$.
A linear pump instead produces an empty-cavity $\pm x$ cat and a continuous
transition into an even-$m$ cat-like state with order parameter
$\langle\hat J_z^2\rangle$.
With cavity loss alone, the circular trajectory emits a photon burst and
approaches a Raman-dark Zeeman pole rather than a finite-field condensate.
In the linear geometry, cavity-bright modes are damped toward the cat-like
branch while a dark spin mode remains undamped.
These are phenomena of that model; the vector-only controls show that
they depend on the channel-fixed tensor coupling.
\end{abstract}

\maketitle

\section{Introduction}
\label{sec:introduction}

The Dicke model describes an ensemble of two-level atoms coupled to a single radiation mode and was originally introduced to explain cooperative spontaneous emission~\cite{Dicke1954}.
In equilibrium, the same Hamiltonian has a superradiant quantum phase transition at a critical collective coupling~\cite{HeppLieb1973,Emary2003}.
The strong-coupling limit of atom--light interaction is difficult to reach, and a steady-state Dicke phase transition was long thought to be elusive.
Through years of effort, the Dicke model has been realized with momentum states playing the role of the two levels, where superradiance of light appears in steady state together with a self-organized density-wave order~\cite{Baumann2010,Baumann2011,Mottl2012,Brennecke2013}.
This paradigm has been extended in several directions, including fermionic systems with Pauli exclusion~\cite{Inouye1999,Moore1999,Keeling2014,Piazza2014b,Chen2014b,Piazza2014a,Chen2015,Zhang2021,Deb2021,Margalit2021,Sanner2021}, strongly correlated bosons in cavity-induced emergent lattices and cavity-mediated long-range interactions~\cite{Klinder2015b,Landig2016,Bhaseen2009,Silver2010,LeBoite2013,Li2013,Bakhtiari2015,Hruby2018,Li2021,Chen2016,Dogra2016,Sundar2016,Chen2022}, and multimode cavities where a continuous $U(1)$ symmetry emerges and its breaking yields gapless modes and supersolid phases~\cite{Kollar2017,Leonard2017}.
These developments underscore the richness of collective light-matter physics~\cite{Ritsch2013,Mivehvar2021}, yet across these platforms the internal degree of freedom is almost universally reduced to an effective two-level spin.

When the cavity is driven and lossy, the problem becomes nonequilibrium~\cite{Klinder2015}: dissipation and continuous pumping select long-time attractors, including coexisting ordered states and persistent oscillations~\cite{Keeling2010,Bhaseen2012}.
Dissipation can also induce new transitions, such as the dissipation-induced superradiant transition observed in strontium~\cite{Song2025}, generate directional flow or limit cycles, and support time-crystalline order---both discrete and continuous---where time-translation symmetry is spontaneously broken~\cite{Gong2018,Kessler2021,Kongkhambut2022}.
These nonequilibrium phases establish cavity QED as a versatile platform, but even in these settings the atomic internal manifold remains a pseudospin-$1/2$.

Open-shell atoms such as erbium, dysprosium, and titanium offer a larger internal manifold.
Erbium and dysprosium combine laser cooling and condensation with ground-state electronic angular momentum $J>1/2$~\cite{Lu2011,Aikawa2012,Chomaz2023}, and titanium has been laser cooled on a high-spin cycling transition~\cite{Eustice2020,Schrott2024,Eustice2025}.
Superradiance has also been observed in a dipolar condensate of $^{168}\mathrm{Er}$, with the atoms in the stretched Zeeman state~\cite{Seo2024}.

For one electric-dipole channel, rotational symmetry fixes the optical response to scalar, vector, and rank-two tensor light shifts.
The coefficients $u_s$, $u_v$, and $u_t$ are set by the angular momenta $(J,J')$ and cannot be tuned independently~\cite{Geremia2006,DeutschJessen2010,LeKien2013}.
The tensor coefficient vanishes at $J=1/2$ and is present for $J\geq1$.
Optical trapping of erbium has resolved both a scalar polarizability and an anisotropic tensor polarizability~\cite{Lepers2014,Becher2018}.
On a large Zeeman manifold the tensor operators are weighted by $m$ or $m^2$, so the differential light shift and the $m$-dependent Raman vertex are comparable to the vector contribution even where $|u_t|$ is smaller than $|u_v|$.
That operator structure cannot be discarded.
We take $^{168}\mathrm{Er}$ with $J=6$ as a concrete product-state example; the same structure applies to any open-shell species on the same electric-dipole channel.
The question is which phenomena this extended Dicke model produces.

Here the optical field is not treated as a prescribed trapping potential.
Instead, $N$ atoms share a dynamical cavity mode whose coherent amplitude responds to the internal spin and may leak from the resonator.
We restrict the atoms to a common spatial orbital and detune the density-wave channel, so the order parameter resides in the Zeeman manifold rather than in a $\lambda/2$ density modulation.
This internal-state truncation places the model on the same broad experimental route as cavity-assisted Raman realizations while isolating the high-spin light-shift structure.

We consider two optical designs obtained from the same effective Hamiltonian.
A circularly polarized pump retains a vector $\hat J_z$ light shift and produces a chiral, raising-type Raman vertex.
A linearly polarized pump cancels that vector shift, activates the pump quadrupole $\hat J_\pm^2$, and couples the vector and tensor Raman operators to orthogonal cavity quadratures.
The two designs are not alternative representations of one model; they are distinct polarization settings in which the tensor interaction has different consequences.
Within each setting, we use $u_t=0$ only as a theoretical comparison that isolates those consequences, not as an independently tunable experimental limit.

The tensor term then gives the two settings different spin order
parameters, a dipole moment for circular pumping and a quadrupole moment
for linear pumping.
The following sections locate the product-state thresholds of each vertex
and the distinct trajectories that appear when only the cavity is lossy.

The remainder of this paper is organized as follows.
Section~\ref{sec:model} introduces the effective polarizability Hamiltonian and the two polarization designs.
Section~\ref{sec:steady-state} analyzes the steady-state phase transitions within product-state mean-field theory.
Section~\ref{sec:stability} examines the linear stability of the stationary states.
Section~\ref{sec:dynamics} presents the cavity-loss dynamics from experimentally preparable initial states.
Section~\ref{sec:conclusion} summarizes the results and the limitations of the model.

\section{Effective model}
\label{sec:model}

\subsection{Polarizability Hamiltonian}
\label{sec:polarizability}

We set $\hbar=1$ and consider a ground manifold of angular momentum $J$ coupled far off resonance to a single odd-parity manifold $J'=J,J\pm1$.
For detuning $\abs{\Delta_e}$ large compared with the excited-state linewidth, adiabatic elimination of the electric-dipole transition gives an interaction within the ground manifold.
Rotational covariance, equivalently the Wigner--Eckart theorem, restricts it to irreducible ranks $0$, $1$, and $2$.
It takes the scalar--vector--tensor form of Ref.~\cite{LeKien2013},
\begin{align}
\hat H_{\mathrm{eff}}
={}&u_s\,\widetilde{\bm E}^{\,*}\!\cdot\widetilde{\bm E}
-\ii u_v
\bigl(\widetilde{\bm E}^{\,*}\!\times\widetilde{\bm E}\bigr)\!\cdot\hat{\bm J}
\notag\\
&+\frac{u_t}{2}\Biggl[
\bigl(\widetilde{\bm E}^{\,*}\!\cdot\hat{\bm J}\bigr)
\bigl(\widetilde{\bm E}\!\cdot\hat{\bm J}\bigr)
+\bigl(\widetilde{\bm E}\!\cdot\hat{\bm J}\bigr)
\bigl(\widetilde{\bm E}^{\,*}\!\cdot\hat{\bm J}\bigr)
\notag\\
&\qquad-\frac{2}{3}
\bigl(\widetilde{\bm E}^{\,*}\!\cdot\widetilde{\bm E}\bigr)
\hat{\bm J}^{\,2}
\Biggr],
\label{eq:effective-H}
\end{align}
with the angular-momentum prefactors of that decomposition absorbed into $u_s$, $u_v$, and $u_t$.
The coefficients contain a common scale set by the reduced dipole matrix
element and optical detuning, whereas their ratios are fixed by angular
momentum and are collected in Appendix~\ref{app:we}.
The $^{168}\mathrm{Er}$ example with $J=6\rightarrow7$ has
$u_s:u_v:u_t=70:15:-2$.
The absolute value of $u_v$ is therefore not fixed to unity.
After absorbing the spin-independent scalar shift into the effective cavity
detuning, we measure all polarizability coefficients in units of $u_v$ and
set $u_v=1$ by convention.
The remaining spin-dependent Hamiltonian is then specified by
$u_t/u_v=-2/15$ for $J\rightarrow J+1$ and
$u_t/u_v=2/11$ for $J\rightarrow J-1$.
These are channel-fixed ratios, not independently tunable couplings.

We take the cavity field to be $\pi$ polarized along $z$, with amplitude $g\hat a$, and denote the pump's circular components by $\Omega_R$ and $\Omega_L$.
A $\pi$ cavity mode and a $\sigma$ pump cannot be collinear because both fields are transverse.
We project onto a common spatial orbital with the density-wave channel off resonance, rather than assuming vanishing two-photon recoil; any residual quadratic recoil shift is absorbed into the diagonal spin Hamiltonian.
Magnetic dipole--dipole interactions, spin-dependent trapping, and atomic dissipation are neglected.
Substituting these fields into Eq.~\eqref{eq:effective-H} gives two
single-atom Hamiltonians, depending on the pump polarization.

\subsection{Circular-pump design}
\label{sec:model-circular}

For a right-circular pump we set $\Omega_L=0$ and choose the phase of $\Omega_R$ to be real.
We use $V_0=\abs{\Omega_R}^2/(2\Delta_e)$,
$U_0=g^2/\Delta_e$, and $\eta_0=g\Omega_R/(2\Delta_e)$.
The two required spin operators are
\begin{align}
\hat K_z^{(c)}&=u_v\hat J_z+u_t\hat J_z^2,
&
\hat K_+^{(c)}&=u_v\hat J_+-u_t\{\hat J_z,\hat J_+\},
\label{eq:circular-operators}
\end{align}
with $\hat K_-^{(c)}=[\hat K_+^{(c)}]^\dagger$.
The single-atom Hamiltonian is
\begin{equation}
\mathcal H_{\mathrm{single}}^{(c)}
=V_0\hat K_z^{(c)}
-U_0u_t\hat a^\dagger\hat a\,\hat J_z^2
+\eta_0\left(
\hat a^\dagger\hat K_+^{(c)}
+\hat a\hat K_-^{(c)}
\right).
\label{eq:circular-single}
\end{equation}
The vector pump shift $u_vV_0\hat J_z$ remains, while the pump-only term proportional to $\hat J_\pm^2$ vanishes.
The Raman vertex is unidirectional in the Zeeman basis:
\begin{align}
\hat K_+^{(c)}\ket{m}
={}&
\left[u_v-u_t(2m+1)\right]
\notag\\
&\times\sqrt{J(J+1)-m(m+1)}
\ket{m+1}.
\label{eq:circular-matrix-element}
\end{align}
Equation~\eqref{eq:circular-matrix-element} is the microscopic starting point for the circular-pump susceptibility and free-energy analysis in Sec.~\ref{sec:circular-results}.

\subsection{Linear-pump design}
\label{sec:model-linear}

For an $x$-polarized pump of real amplitude $\Omega$, the circular components are
$\Omega_R=-\Omega/\sqrt2$ and $\Omega_L=\Omega/\sqrt2$.
Their equal intensities cancel the vector $\hat J_z$ pump shift, whereas
$\Omega_L^*\Omega_R=-\abs{\Omega}^2/2$ activates the pump quadrupole.
We use $V_0=\Omega^2/(2\Delta_e)$, $U_0=g^2/\Delta_e$, and
$\eta_0=g\Omega/(2\sqrt2\,\Delta_e)$, and define
\begin{align}
\hat K_z^{(l)}
&=u_t\left[
\hat J_z^2-2(\hat J_x^2-\hat J_y^2)
\right],
\notag\\
\hat K^{(l)}
&=-2\ii u_v\hat J_y
+2u_t\{\hat J_z,\hat J_x\},
\label{eq:linear-operators}
\end{align}
The single-atom Hamiltonian is
\begin{equation}
\mathcal H_{\mathrm{single}}^{(l)}
=V_0\hat K_z^{(l)}
-\frac{U_0u_t}{2}\hat a^\dagger\hat a\,\hat J_z^2
+\eta_0\left(
\hat a^\dagger\hat K^{(l)}
+\hat a[\hat K^{(l)}]^\dagger
\right).
\label{eq:linear-single}
\end{equation}
The vector part of $\hat K^{(l)}$ couples to $\operatorname{Im}\alpha$, while its tensor part couples to $\operatorname{Re}\alpha$.
Unlike Eq.~\eqref{eq:circular-single}, this design contains bidirectional Raman transitions together with the pump-induced $\hat J_\pm^2$ term encoded in $\hat K_z^{(l)}$.
Equations~\eqref{eq:circular-single} and~\eqref{eq:linear-single} therefore describe two distinct polarization settings derived from the same scalar--vector--tensor interaction.

\subsection{Many-body mean-field reduction}
\label{sec:many-body-reduction}

We now embed either single-atom Hamiltonian in a system of $N$ atoms coupled
to the same cavity mode:
\begin{equation}
\hat H=\delta_c\hat a^\dagger\hat a+
\sum_{i=1}^{N}\mathcal H_{\mathrm{single}}^{(i)},
\label{eq:many-body-general}
\end{equation}
where $\mathcal H_{\mathrm{single}}$ is chosen as
Eq.~\eqref{eq:circular-single} or Eq.~\eqref{eq:linear-single}.
Here $\delta_c$ includes the collective scalar cavity shift, and
spin-independent constants have been dropped.
The remainder of this work concerns the mean-field many-body problem.
Specifically, we use a coherent cavity field
$\alpha=\avg{\hat a}/\sqrt N$ and a common normalized
$(2J+1)$-component spinor for all atoms, with collective Raman coupling
$\eta=\sqrt N\,\eta_0$.
This product-state closure retains the complete Zeeman manifold but neglects
finite-$N$ correlations.

We henceforth use the normalization $u_v=1$.
Numerical illustrations use $J=6$, with
$u_t=-2/15$ in the circular $J\rightarrow J+1$ design and
$u_t=2/11$ in the linear $J\rightarrow J-1$ design.
Working values of $V_0$, $NU_0$, $\delta_c$, and $\kappa$ are those of the
corresponding figures.
In each polarization setting we also evaluate the formal limit $u_t=0$ while holding the remaining design fixed.
This comparison diagnoses the role of the tensor block and is not an additional experimental control assumed by the model.

\section{Steady-state phase transitions}
\label{sec:steady-state}

Within the many-body mean-field description introduced above and with atomic
dissipation omitted, each optical design has a product-state free energy whose
minima determine the steady states.
The two polarizations are analyzed separately.

\subsection{Circular pump}
\label{sec:circular-results}

\begin{figure}[t]
\centering
\includegraphics[width=\columnwidth]{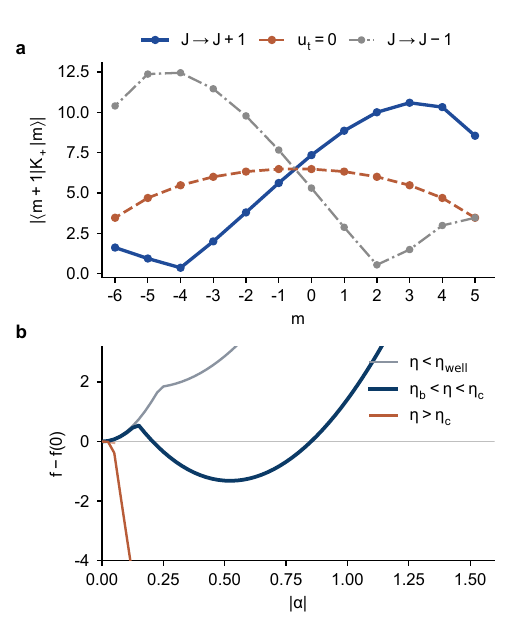}
\caption{Circular-pump mechanism for $J=6$, $V_0=0.2$, and $NU_0=5$.
(a) Raman matrix elements for the working $J\rightarrow J+1$ channel, the
$u_t=0$ control, and the reversed $J\rightarrow J-1$ channel.
(b) Free-energy cuts at $\delta_c=10$.
The thick curve lies between the first-order transition $\eta_b$ and the
empty-cavity instability $\eta_c$: its finite-field minimum is globally
favored while the origin remains locally stable.}
\label{fig:c-mechanism}
\end{figure}

We first consider the red-detuned $J\rightarrow J+1$ channel, for which
$u_v=1$ and $u_t=-2/15$ in our normalization.
At $\alpha=0$, the single-spin energy
$E(m)=V_0(u_vm+u_tm^2)$ is minimized by $\ket{m=-J}$.
The tensor term nevertheless makes the Raman coupling strongly dependent on
$m$: as shown in Fig.~\ref{fig:c-mechanism}(a),
$\abs{\bra{m+1}\hat K_+^{(c)}\ket m}$ is weak at $m=-J$ and largest at
positive $m$.
A finite cavity field can therefore lower its energy by transferring
population away from the empty-cavity state and into the positive-$m$ Raman
window.
This selection is absent from the vector-only control, whose matrix element
is largest near the equator.

To distinguish the onset of a finite-field minimum from the loss of stability
of the empty cavity, we minimize the spin energy at fixed
$r=\abs{\alpha}$.
Near the origin, the product-state free energy is obtained by expanding
around $\ket{\psi_0}=\ket{m=-J}$:
\begin{align}
f(r)-f(0)
={}&\left(\delta_{\mathrm{eff}}+\eta^2\chi_c\right)r^2
+\mathcal O(r^4),
\notag\\
\chi_c
={}&\sum_{n\ne0}
\frac{\abs{\bra n(\hat K_+^{(c)}+\hat K_-^{(c)})
\ket{\psi_0}}^2}{E_0-E_n},
\label{eq:c-susceptibility}
\end{align}
where
$\delta_{\mathrm{eff}}
=\delta_c-NU_0u_t\avg{\hat J_z^2}_{\psi_0}$.
Since $\chi_c<0$, the empty-cavity branch loses local stability at
\begin{equation}
\eta_c=\sqrt{\frac{\delta_{\mathrm{eff}}}{\abs{\chi_c}}}.
\label{eq:c-critical}
\end{equation}

Equations~\eqref{eq:c-susceptibility} and~\eqref{eq:c-critical} expand the free energy perturbatively about $r=0$.
They fix the curvature of the empty-cavity branch and locate only its local instability $\eta_c$.
For circular pumping, the tensor term makes the Raman vertex in Eq.~\eqref{eq:circular-matrix-element} depend on $m$.
That nonlinear structure produces the first-order transition shown in Fig.~\ref{fig:c-mechanism}(b).
A finite-$r$ local minimum first appears at $\eta_{\mathrm{well}}$ and becomes degenerate with the origin at $\eta_b$, while the origin remains locally stable until $\eta_c$.
Numerical minimization of the free energy yields $\eta_{\mathrm{well}}<\eta_b<\eta_c$.
The jump occurs because $\ket{m=-J}$ lies at the weak-coupling end of Fig.~\ref{fig:c-mechanism}(a): a small displacement in $r$ admixes only the nearby Zeeman state and gains little energy.
A finite cavity amplitude transfers population into the positive-$m$ window, where $\abs{\bra{m+1}\hat K_+^{(c)}\ket{m}}$ is largest, and the free energy of that rearranged state crosses $f(0)$ at $\eta_b$ before the curvature at the origin changes sign.
Thus, for $\eta_b<\eta<\eta_c$, the finite-field state is the global minimum although the empty-cavity branch remains a local minimum.
When $u_t$ is set to zero, the vertex is no longer suppressed at $m=-J$, and the three scales merge into a single continuous transition.

\begin{figure*}[t]
\centering
\includegraphics[width=\textwidth]{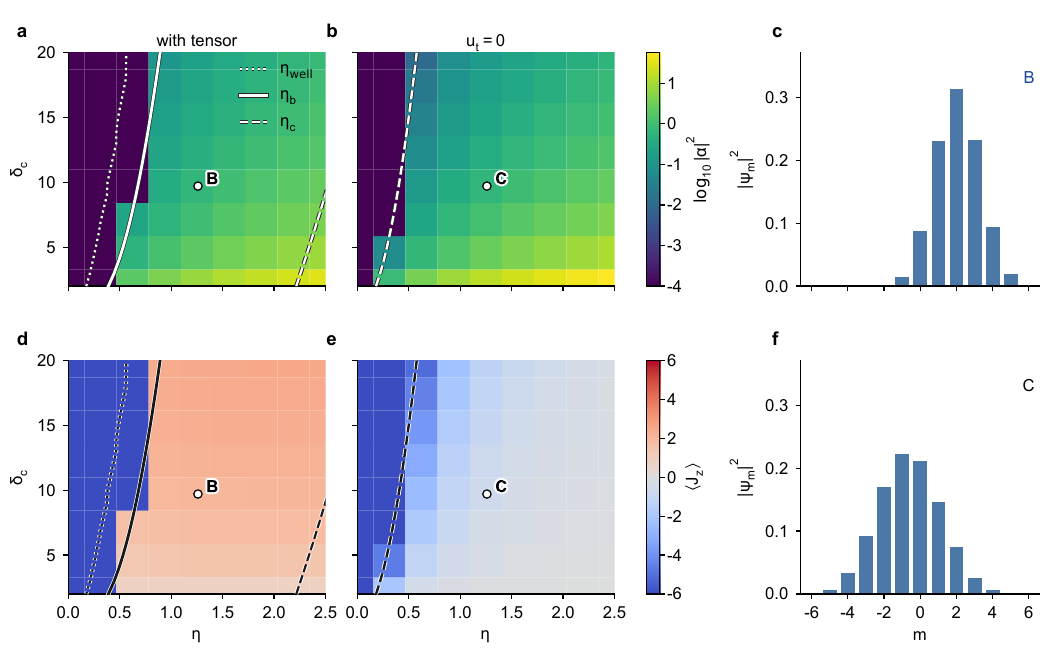}
\caption{Circular-pump steady-state phase diagram and Zeeman populations at
$J=6$, $V_0=0.2$, $NU_0=5$, and $\kappa=0$.
The upper row shows $\log_{10}\abs{\alpha}^2$ and the lower row
$\avg{\hat J_z}$; the left (right) heatmap column includes the tensor term
(sets $u_t=0$).
Dotted, solid, and dashed curves denote the appearance of a finite-field
local minimum, the first-order transition $\eta_b$, and the loss of local
stability of the empty-cavity branch $\eta_c$, respectively.
For $u_t=0$ they coincide at a continuous transition.
The right column shows the distributions at the marked points:
(c) point B;
(f) point C at the same $(\eta,\delta_c)$ as B with $u_t=0$.}
\label{fig:c-phase}
\end{figure*}

The phase diagram in Fig.~\ref{fig:c-phase} presents the minima of the same
free energy over $(\eta,\delta_c)$.
With the tensor term, both the photon number and the magnetization change
discontinuously at $\eta_b$, well before the empty-cavity branch loses local
stability at $\eta_c$.
The condensed state occupies positive $m$ and hence has
$\avg{\hat J_z}>0$, opposite to the empty state
$\ket{m=-J}$.
For the $u_t=0$ control, photon condensation still occurs, but the transition
is continuous and the magnetization at the same working point has the
opposite sign.
The tensor term therefore changes both the order of the transition and the
Zeeman sector selected by the cavity; it does not merely change the photon
number.

The Zeeman distributions make this change explicit.
At point B in Fig.~\ref{fig:c-phase}(c), the condensed packet peaks near
$m\simeq2$, in the strong-coupling window of
Fig.~\ref{fig:c-mechanism}(a).
At the same $(\eta,\delta_c)$ as B but with $u_t=0$, point C in
Fig.~\ref{fig:c-phase}(f) instead produces a broad packet around
$m\simeq-1$.
Consequently, $\avg{\hat J_z}$ is the natural spin readout for the circular
geometry.

\subsection{Linear pump}
\label{sec:linear-results}

The linear-pump geometry has a qualitatively different empty-cavity state.
For the red-detuned $J\rightarrow J-1$ channel,
$u_t/u_v=2/11$, the operator $\hat K_z^{(l)}$ can be written, up to a
constant, as
\begin{equation}
\hat K_z^{(l)}
=u_t\left[2J(J+1)-4\hat J_x^2-\hat J_z^2\right].
\label{eq:l-empty-H}
\end{equation}
It therefore favors the two semiclassical orientations $\pm x$.
At finite $J$, the exact ground state is their even-$m$ superposition: an
even $\pm x$ cat with
$\avg{\hat{\bm J}}\simeq0$,
a quadratic moment of order unity, and a population maximum at $m=0$.
The odd partner is nearly degenerate, with a cat splitting far below the
remaining excitation gaps.

This small cat splitting does not set the transition threshold.
The cat partner is dark to both Raman quadratures, and the response is instead
controlled by the next even/odd excited pair, whose gap remains finite.
Writing $\alpha=x+\ii p$, the expansion of the free energy is
\begin{align}
f(x,p)-f(0)
={}&
\begin{pmatrix}x&p\end{pmatrix}
\left[\delta_{\mathrm{eff}}\mathbb I+\eta^2\bm\chi\right]
\begin{pmatrix}x\\p\end{pmatrix}
+\mathcal O(\abs{\alpha}^4),
\notag\\
\chi_{ab}
={}&\sum_{n\ne0}
\frac{\bra0\hat V_a\ket n\bra n\hat V_b\ket0}{E_0-E_n},
\notag\\[-2pt]
&\hspace{4em}a,b\in\{x,p\},
\label{eq:l-susceptibility}
\end{align}
where $\hat V_x$ and $\hat V_p$ are the Hermitian Raman vertices coupled to
$x$ and $p$:
$\hat V_x=\hat K^{(l)}+[\hat K^{(l)}]^\dagger$ and
$\hat V_p=\ii([\hat K^{(l)}]^\dagger-\hat K^{(l)})$.
At the empty-cavity cat, $\chi_{xp}=0$ and
$\abs{\chi_{xx}}\gg\abs{\chi_{pp}}$, so the real tensor
quadrature is soft.
The continuous transition occurs at
\begin{equation}
\eta_c=
\sqrt{\frac{\delta_{\mathrm{eff}}}
{\abs{\lambda_{\min}(\bm\chi)}}}.
\label{eq:l-critical}
\end{equation}

\begin{figure}[t]
\centering
\includegraphics[width=\columnwidth]{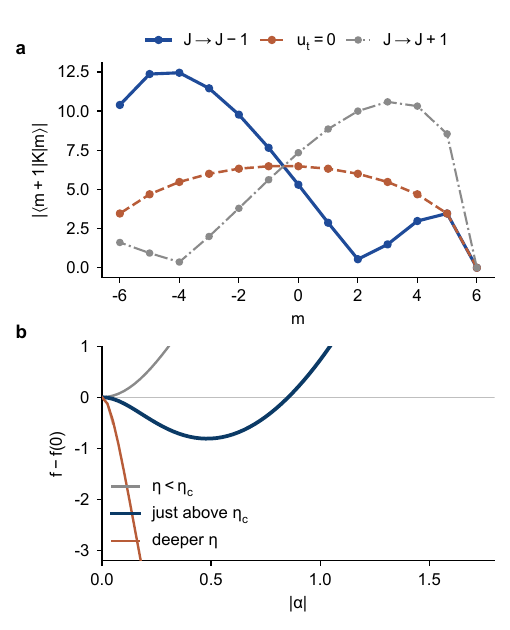}
\caption{Linear-pump mechanism for $J=6$, $V_0=0.2$, and $NU_0=5$.
(a) Raman weight
$\abs{\bra{m+1}\hat K^{(l)}\ket m}$ for the working
$J\rightarrow J-1$ channel and two controls.
The working channel enhances the polar regions and has nodes near
$m=\pm2$.
(b) Free-energy cuts at $\delta_c=20$ below, just above, and deeper beyond
$\eta_c$.
The finite-field minimum appears continuously when the origin loses
stability.}
\label{fig:l-mechanism}
\end{figure}

Figure~\ref{fig:l-mechanism}(a) connects this response to the Zeeman basis.
The working-channel Raman weight is enhanced toward both poles and develops
nodes near $m=\pm2$.
Once the real cavity quadrature becomes finite, the tensor vertex transfers
weight from the equatorial empty cat toward large $\abs m$.
Figure~\ref{fig:l-mechanism}(b) shows that the finite-field minimum emerges
continuously as the curvature at the origin changes sign.
Unlike the circular case, the appearance of this minimum, its crossing with
$f(0)$, and the instability of the origin all occur at $\eta_c$.

\begin{figure*}[t]
\centering
\includegraphics[width=\textwidth]{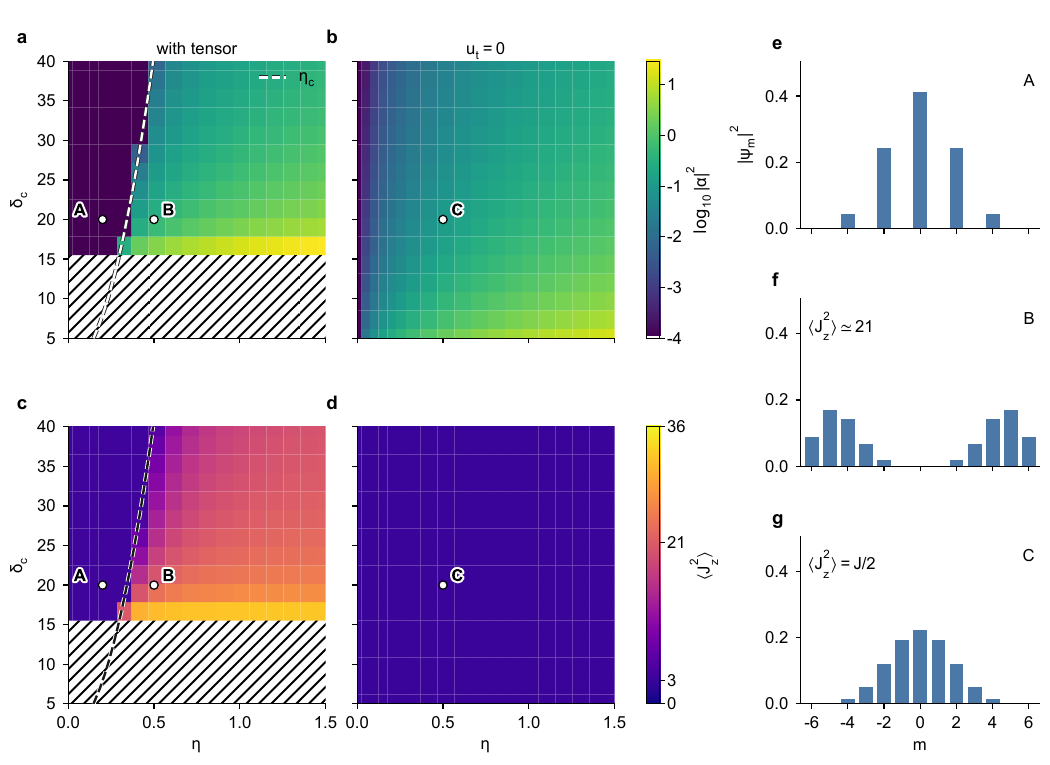}
\caption{Linear-pump steady-state phase diagram and Zeeman populations at
$J=6$, $V_0=0.2$, $NU_0=5$, and $\kappa=0$.
The upper row shows $\log_{10}\abs{\alpha}^2$ and the lower row
$\avg{\hat J_z^2}$; the left (right) heatmap column includes the tensor term
(sets $u_t=0$).
The dashed curve is the continuous transition.
Hatching marks $\delta_c\lesssim16.4$ in Eq.~\eqref{eq:l-bounded}, where the product-state free energy is unbounded.
The right column shows the Zeeman populations at the marked points:
(e) empty-cavity point A, an even-$m$ $\pm x$ cat peaked at $m=0$;
(f) condensed point B, an even-$m$ cat-like state with weight near
$\abs m\simeq5$ and $\avg{\hat J_z^2}\simeq21$;
(g) vector-only point C, a $y$-polarized coherent state with
$\avg{\hat J_z^2}=J/2$.}
\label{fig:l-phase}
\end{figure*}

The full phase diagram is shown in Fig.~\ref{fig:l-phase}.
In the bounded region, the photon field turns on continuously across
Eq.~\eqref{eq:l-critical}, while
$\avg{\hat J_z^2}$ grows from its empty-cat value toward
$\mathcal O(J^2)$.
The first moment is not a useful order parameter because
$\avg{\hat{\bm J}}\simeq0$ on both sides.
The hatched region is where this free energy is unbounded.
Collecting the dispersive term of Eq.~\eqref{eq:linear-single} over $N$ atoms
shifts the cavity detuning by $-\frac12 NU_0 u_t\avg{\hat J_z^2}$.
On the working channel $u_t>0$, that shift can cancel $\delta_c$ once
$\avg{\hat J_z^2}$ approaches $J^2$, so the free energy stays bounded only for
\begin{equation}
\delta_c\gtrsim\tfrac12 N|U_0|\,u_t J^2.
\label{eq:l-bounded}
\end{equation}
The cuts and the phase diagram use $\delta_c=20$, above this line.
The red-detuned circular design has $NU_0 u_t<0$, so its dispersive shift
stiffens the cavity oscillator and the free energy remains bounded for every
$\delta_c>0$.
Without the tensor term, $\hat K_z^{(l)}$ vanishes, the origin is unstable
for every $\eta>0$, and the minimum is a $y$-polarized coherent state with
$\avg{\hat J_z^2}=J/2$.

Figure~\ref{fig:l-phase} identifies the states on both sides of the
transition.
The empty state A in Fig.~\ref{fig:l-phase}(e) is the even-$m$ $\pm x$ cat,
concentrated around $m=0$.
In the condensed state B of Fig.~\ref{fig:l-phase}(f), only even $m$ remain
occupied but the distribution moves to $\abs m\simeq4$--$5$, with
$\avg{\hat J_z^2}\simeq21$ and
$\avg{\hat{\bm J}}\simeq0$.
We therefore refer to it as a cat-like state rather than a two-component cat:
it differs from the empty $\pm x$ cat in both its Zeeman distribution and its
quadratic moment.
The vector-only control C in Fig.~\ref{fig:l-phase}(g) is instead the
$y$-polarized coherent state with $\avg{\hat J_z^2}=J/2$.
The contrast between B and C shows that the large-$\abs m$ occupation is a
tensor effect, not a generic consequence of cavity occupation.

The circular and linear results therefore isolate two different tensor
mechanisms within the same microscopic polarizability Hamiltonian.
Circular pumping produces a first-order transition before the empty-cavity
branch becomes unstable and selects the sign of
$\avg{\hat J_z}$.
Linear pumping instead produces a finite continuous threshold and a
cat-like condensate whose order parameter is
$\avg{\hat J_z^2}$.
Channel reversal and blue-detuned counterparts retain the same operator
logic; the four-cell classification is in Appendix~\ref{app:channels}.

\section{Linear stability}
\label{sec:stability}

The free-energy analysis of Sec.~\ref{sec:steady-state} locates product-state
minima at $\kappa=0$.
Whether those minima are dynamically stable, and which of them survive at
finite cavity loss, is decided by the product-state flow of a common spinor
$\psi$ together with a coherent cavity amplitude $\alpha$.
The two polarizations are linearized separately.

\subsection{Circular pump}
\label{sec:stability-circular}

The spinor evolves under Eq.~\eqref{eq:circular-single} after the collective
replacements $\eta_0\rightarrow\eta$ and $U_0\rightarrow NU_0$:
\begin{equation}
\begin{aligned}
\ii\dot\psi
={}&
\bigl[
V_0\hat K_z^{(c)}
-NU_0u_t\abs{\alpha}^2\hat J_z^2
+\eta\bigl(\alpha^*\hat K_+^{(c)}+\alpha\hat K_-^{(c)}\bigr)
\bigr]\psi,
\\
\ii\dot\alpha
={}&
\bigl(\delta_c-NU_0u_t\avg{\hat J_z^2}-\ii\kappa\bigr)\alpha
+\eta\avg{\hat K_+^{(c)}}.
\end{aligned}
\label{eq:circular-flow}
\end{equation}
At $\kappa=0$ this flow conserves the energy, so a free-energy minimum is a
fixed point rather than an attractor.
For real variables
$y=(\operatorname{Re}\alpha,\operatorname{Im}\alpha,
\operatorname{Re}\psi,\operatorname{Im}\psi)$, a perturbation about a fixed
point obeys
\begin{equation}
\delta\dot y=\mathsf M\,\delta y .
\label{eq:linearized-flow}
\end{equation}
Gauge zeros associated with the overall phase of $\psi$ are discarded,
equivalently replacing the spin Hamiltonian by
$H(\alpha)-\avg{H(\alpha)}$ so that a ground-state pair is a true
fixed point of Eq.~\eqref{eq:circular-flow}.

At $\kappa=0$, every free-energy minimum is an elliptic center
(Table~\ref{tab:c-jac}):
$\max\operatorname{Re}\lambda\simeq0$ with a nonzero imaginary part.
Point B lies between the first-order transition and the empty-cavity
instability, so empty and condensed centers coexist.

\begin{table}[t]
\caption{Circular-pump linearization at $\kappa=0$ for the markers of
Fig.~\ref{fig:c-phase}.
$(\eta,\delta_c)$ is $(0.32,10)$ at A and $(1.26,9.714)$ at B and C;
C sets $u_t=0$.
A full-space attractor would require $\max\operatorname{Re}\lambda<0$.
At $\kappa=1$ the condensed seed is not a fixed point, so it has no row here.}
\label{tab:c-jac}
\centering
\small
\begin{tabular}{llccccl}
\hline
pt & branch & $\abs{\alpha}^2$ & $\avg{\hat J_z}$ & $\max\operatorname{Re}\lambda$ & $\min\operatorname{Re}\lambda$ & class \\
\hline
A & both & $0$ & $-6.00$ & $\sim0$ & $\sim0$ & elliptic \\
B & empty & $0$ & $-6.00$ & $\sim0$ & $\sim0$ & elliptic \\
B & cond. & $0.620$ & $+2.03$ & $\sim0$ & $\sim0$ & elliptic \\
C & cond. & $0.599$ & $-0.61$ & $\sim0$ & $\sim0$ & elliptic \\
\hline
\end{tabular}
\end{table}

\begin{figure}[t]
\centering
\includegraphics[width=\columnwidth]{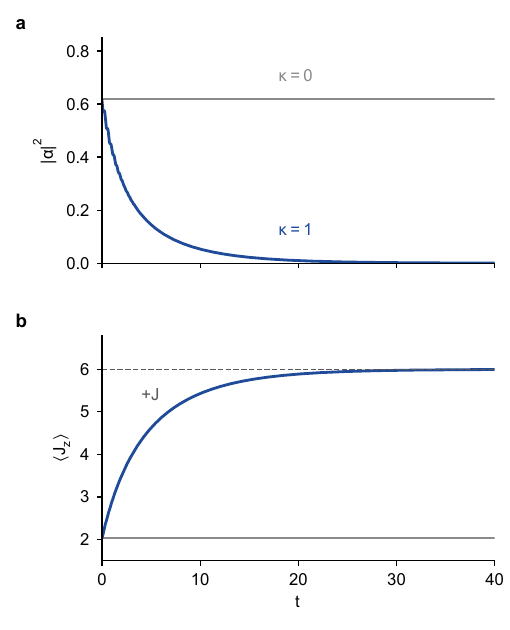}
\caption{Circular condensed state at point B.
(a)~Photon number and (b)~$\avg{\hat J_z}$ from the condensate itself.
Gray ($\kappa=0$): the trajectory remains in the condensed well.
Blue ($\kappa=1$): photons leak and the spin approaches $+J$.}
\label{fig:c-kick}
\end{figure}

Figure~\ref{fig:c-kick} starts from the condensed center at point B
itself (Table~\ref{tab:c-jac}), without a kick.
At $\kappa=0$ the photon number remains at $\abs{\alpha}^2\simeq0.62$
and $\avg{\hat J_z}$ at the condensed value $\simeq+2$.
The same initial state with $\kappa=1$ empties the cavity:
$\abs{\alpha}^2\to0$ and the spin approaches the dark pole $+J$.
The equatorial spin-coherent trajectory of Sec.~\ref{sec:dynamics} lies on a
different energy surface, $E\simeq-0.08$ versus
$E_{\mathrm{GS}}\simeq-8.12$, and cannot reach this center.

Cavity loss does not turn these centers into attractors.
Iteration from the empty branch remains at $m=-J$ below its instability,
whereas continuation of the finite-field free-energy minimum does not
yield a converged cavity-condensed fixed point in the parameter regime
studied.

\subsection{Linear pump}
\label{sec:stability-linear}

The corresponding flow from Eq.~\eqref{eq:linear-single} is
\begin{equation}
\begin{aligned}
\ii\dot\psi
={}&
\bigl[
V_0\hat K_z^{(l)}
-\tfrac12 NU_0u_t\abs{\alpha}^2\hat J_z^2
\\
&\quad
+\eta\bigl(\alpha^*\hat K^{(l)}+\alpha[\hat K^{(l)}]^\dagger\bigr)
\bigr]\psi,
\\
\ii\dot\alpha
={}&
\bigl(\delta_c-\tfrac12 NU_0u_t\avg{\hat J_z^2}-\ii\kappa\bigr)\alpha
+\eta\avg{\hat K^{(l)}}.
\end{aligned}
\label{eq:linear-flow}
\end{equation}
Perturbations are classified by the same linearization
Eq.~\eqref{eq:linearized-flow}.
At $\kappa=0$ the continuous transition has no coexistence window
(Table~\ref{tab:l-jac}):
both seeds give the empty cat at A and the cat-like state at B.
Every free-energy minimum is again an elliptic center.
The empty cat remains a self-consistent $\alpha=0$ point past $\eta_c$ but
is dynamically unstable at B, $\max\operatorname{Re}\lambda\simeq1.06$.

\begin{table}[t]
\caption{Linear-pump linearization at $\kappa=0$ for the markers of
Fig.~\ref{fig:l-phase}.
$(\eta,\delta_c)$ is $(0.20,20)$ at A and $(0.50,20)$ at B and C;
C sets $u_t=0$.
A full-space attractor would require $\max\operatorname{Re}\lambda<0$.}
\label{tab:l-jac}
\centering
\small
\begin{tabular}{llccccl}
\hline
pt & branch & $\abs{\alpha}^2$ & $\avg{\hat J_z^2}$ & $\max\operatorname{Re}\lambda$ & $\min\operatorname{Re}\lambda$ & class \\
\hline
A & both & $0$ & $3.51$ & $\sim0$ & $\sim0$ & elliptic \\
B & empty & $0$ & $3.51$ & $1.06$ & $-1.06$ & unstable \\
B & cond. & $0.330$ & $21.0$ & $\sim0$ & $\sim0$ & elliptic \\
C & cond. & $0.090$ & $3.00$ & $\sim0$ & $\sim0$ & elliptic \\
\hline
\end{tabular}
\end{table}

Cavity loss does not turn these centers into full-space attractors.
On the bounded side of the phase diagram the cat-like branch still exists at
$\kappa=1$ (Table~\ref{tab:l-jac-k1}).
Every cat-like solution is cavity-damped:
$\max\operatorname{Re}\lambda\simeq0$ with two modes below
$-0.05$.
The empty cat at B remains unstable, $\max\operatorname{Re}\lambda\simeq0.99$.
At the cat-like point B the active spectrum sits on the imaginary axis when
$\kappa=0$; at $\kappa=1$ two bright pairs move to
$\operatorname{Re}\lambda\simeq-0.81$ and $-0.19$, while the remainder stays
at $\operatorname{Re}\lambda\simeq0$.
A Raman-dark spin mode therefore remains undamped; the finite-field
solution is attractive along cavity-bright directions but is not an
attractor in the full spin--cavity phase space.
Its imaginary eigenvalue produces undamped oscillations, displayed later in
Fig.~\ref{fig:l-dynamics}(c).

\begin{table}[t]
\caption{Linear-pump linearization at $\kappa=1$ for the same markers and
$(\eta,\delta_c)$ as Table~\ref{tab:l-jac}.
A full-space attractor would require $\max\operatorname{Re}\lambda<0$.}
\label{tab:l-jac-k1}
\centering
\small
\begin{tabular}{llccccl}
\hline
pt & branch & $\abs{\alpha}^2$ & $\avg{\hat J_z^2}$ & $\max\operatorname{Re}\lambda$ & $\min\operatorname{Re}\lambda$ & class \\
\hline
A & both & $0$ & $3.51$ & $\sim0$ & $-0.99$ & cavity damped \\
B & empty & $0$ & $3.51$ & $0.99$ & $-1.13$ & unstable \\
B & cond. & $0.324$ & $20.9$ & $\sim0$ & $-0.81$ & cavity damped \\
C & cond. & $0.090$ & $3.00$ & $\sim0$ & $-1.00$ & cavity damped \\
\hline
\end{tabular}
\end{table}

\begin{figure}[t]
\centering
\includegraphics[width=\columnwidth]{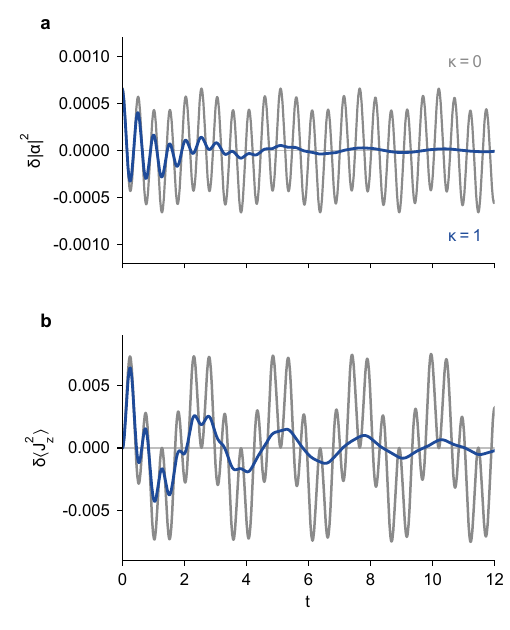}
\caption{Linear cat-like state at point B after a $10^{-3}$ radial
kick of $\alpha$.
Deviations of (a)~the photon number and (b)~$\avg{\hat J_z^2}$ from
the cat-like condensate.
Gray ($\kappa=0$): undamped oscillation.
Blue ($\kappa=1$): return to the cat-like values.}
\label{fig:l-bright}
\end{figure}

Figure~\ref{fig:l-bright} applies a $10^{-3}$ radial kick of $\alpha$
at the cat-like point B.
Deviations of the photon number and of $\avg{\hat J_z^2}$ from the
cat-like values ($\abs{\alpha}^2\simeq0.32$, $\avg{\hat J_z^2}\simeq21$)
decay back to zero at $\kappa=1$ and oscillate without decay at
$\kappa=0$.
The Raman-dark kick remains undamped even at $\kappa=1$ and is shown in
Fig.~\ref{fig:l-dynamics}(c), not here.

\section{Cavity-loss dynamics}
\label{sec:dynamics}

The classification of Sec.~\ref{sec:stability} applies to self-consistent
fixed points.
It does not by itself determine which state is reached from a preparable
initial condition when photons leak from the cavity.
We therefore evolve Eqs.~\eqref{eq:circular-flow} and~\eqref{eq:linear-flow}
from experimentally
simpler starts: an equatorial spin-coherent state for circular pumping and
a $+y$ coherent state for linear pumping, each with a negligible cavity seed.

\subsection{Circular pump: decay into a dark pole}
\label{sec:circular-dynamics}

\begin{figure*}[t]
\centering
\includegraphics[width=\textwidth]{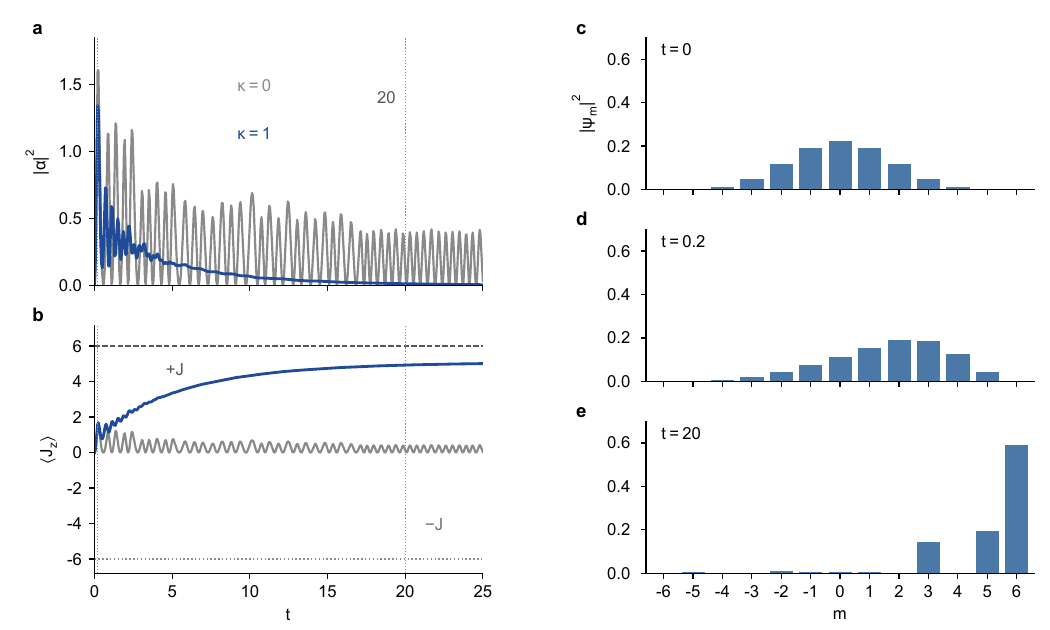}
\caption{Circular-pump dynamics at point B from an equatorial spin-coherent
state.
(a) Photon number and (b) magnetization for $\kappa=0$ (gray) and
$\kappa=1$ (blue).
Without loss the trajectory oscillates; with loss it emits a photon burst,
after which $\avg{\hat J_z}$ approaches the dark pole $+J$.
(c)--(e) Zeeman populations along the $\kappa=1$ trajectory at
$t=0$, $0.2$, and $20$, respectively.}
\label{fig:c-dynamics}
\end{figure*}

We initialize the circular model at point B with an equatorial spin-coherent
state and a negligible cavity seed.
As shown by the gray curves in Fig.~\ref{fig:c-dynamics}, for $\kappa=0$ the
photon number and $\avg{\hat J_z}$ remain locked in persistent oscillations:
the magnetization stays well below $J$, and the orbit cannot
relax to a minimum of the free energy.

Cavity loss changes this trajectory qualitatively.
A finite loss rate produces a photon burst after which the cavity population
decays to zero.
During the burst, the chiral Raman vertex transfers population only toward
larger $m$.
The magnetization consequently rises from zero toward the upper pole, while
the distribution accumulates at $m=5$ and $6$.
The late state is therefore neither the empty-cavity ground state
$\ket{m=-J}$ nor the finite-field minimum of Sec.~\ref{sec:circular-results};
it approaches the Raman-dark pole $\ket{m=+J}$ after the photons have leaked
away.
That outcome is consistent with Sec.~\ref{sec:stability-circular}: at finite $\kappa$
there is no converged cavity-condensed fixed point, so cavity loss does not
select the lower free-energy well of the closed system.
It instead removes the field that would return population from the
Raman-dark upper pole.

\subsection{Linear pump: bright damping and a dark spin mode}
\label{sec:linear-dynamics}

\begin{figure*}[t]
\centering
\includegraphics[width=\textwidth]{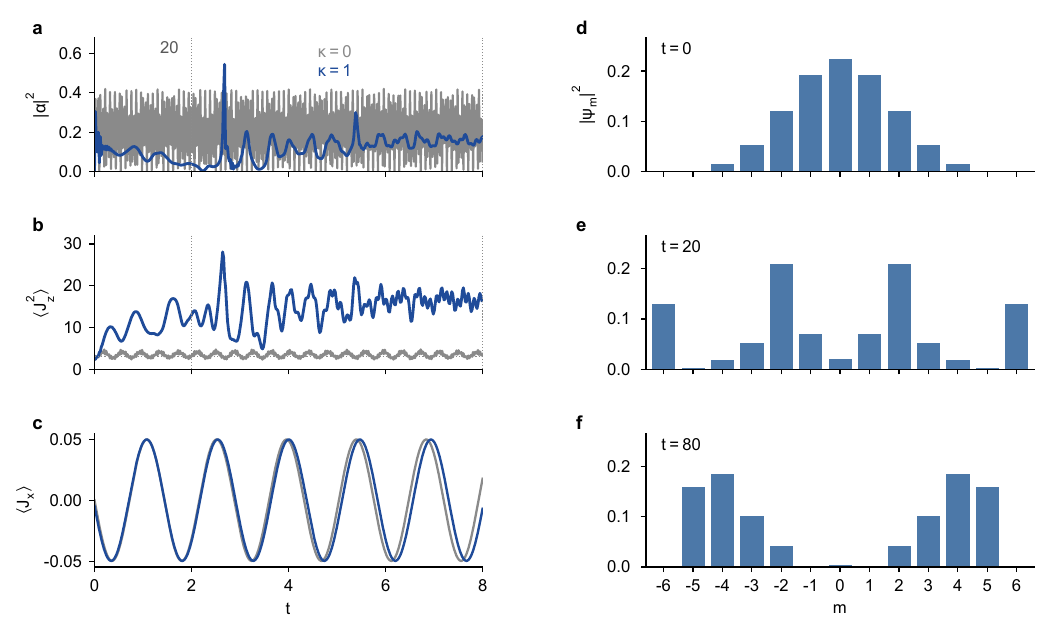}
\caption{Linear-pump dynamics at point B.
(a) Photon number and (b) $\avg{\hat J_z^2}$ from a $+y$ spin-coherent
state for $\kappa=0$ (gray) and $\kappa=1$ (blue).
Only the lossy trajectory moves along the bright direction toward the
cat-like branch.
(c) A kick along a Raman-dark mode produces undamped oscillations of
$\avg{\hat J_x}$ at both loss rates.
(d)--(f) Zeeman populations on the $\kappa=1$ trajectory at
$t=0$, $20$, and $80$, respectively.}
\label{fig:l-dynamics}
\end{figure*}

The linear geometry retains a finite-field self-consistent branch when
$\kappa>0$ on the bounded side of the phase diagram.
The continuous threshold is nearly unchanged at the loss rate used in
Fig.~\ref{fig:l-dynamics}.
We test its accessibility by starting at point B from a $+y$ spin-coherent
state with a negligible cavity seed.
For $\kappa=0$, the photon number continues to oscillate and
$\avg{\hat J_z^2}$ stays near its coherent-state value $J/2$.
For finite cavity loss, the optical oscillations are damped toward a finite
late-time population, while $\avg{\hat J_z^2}$ rises toward the large value of
the cat-like branch [Figs.~\ref{fig:l-dynamics}(a) and
\ref{fig:l-dynamics}(b)].

The Zeeman snapshots show how this change occurs.
The initial binomial distribution is centered at $m=0$.
At intermediate times, even-$m$ components near $\abs m=2$ and $6$ have
acquired substantial weight; later, the distribution is concentrated near
$\abs m=4$--$5$ and the central component is almost empty.
The trajectory is moving toward the cat-like sector identified in
Fig.~\ref{fig:l-phase}(f), although residual oscillations mean that
the latest snapshot is not identical to the stationary minimum.

The incomplete relaxation follows from the linearized spectrum of
Sec.~\ref{sec:stability-linear}: cavity-bright directions acquire negative real
parts, while Raman-dark spin modes remain on the imaginary axis.
Figure~\ref{fig:l-dynamics}(c) excites one such mode:
$\avg{\hat J_x}$ oscillates with essentially the
same amplitude with and without cavity loss.

The two trajectories demonstrate that cavity loss cannot be inferred by simply
minimizing the closed-system free energy.
In the circular geometry it removes the field and leaves the atoms near a
dark Zeeman pole.
In the linear geometry it damps the bright component toward a finite-field
cat-like branch while leaving a dark oscillatory direction.
Atomic relaxation would be required to turn the latter into a full-space
attractor; atomic baths and possible limit-cycle dynamics are outside the
present model.

\section{Conclusion}
\label{sec:conclusion}

Open-shell atoms coupled to a shared cavity mode define an extended Dicke
model through their scalar--vector--tensor polarizability; the tensor
coupling is fixed by the electric-dipole channel.
Empty-cavity Raman susceptibilities locate the local stability thresholds of
the empty branch in both pump geometries.
For circular pumping at $\kappa=0$, a first-order free-energy window
$\eta_b<\eta_c$ appears before the empty-cavity branch loses local
stability, and the condensed Zeeman packet occupies the positive-$m$ window
selected by $\hat K_+^{(c)}$ matrix elements.
For linear pumping, the same empty-cavity threshold is a continuous
transition into an even-$m$ cat-like condensate whose order parameter is
$\avg{\hat J_z^2}$.
Both ordered states disappear when the tensor block is removed within the
same polarization setting, showing that they depend on the tensor coupling
of the model; this comparison is a theoretical diagnostic.
With finite cavity loss, cavity-bright directions of the linear-pump
cat-like state are damped while a Raman-dark spin mode remains undamped;
the circular trajectory emits a photon burst and approaches the dark pole
$m=+J$.

The preceding discussion yields an extended Dicke model, and the new
phenomena arise from that model.
Coexisting empty and finite-field solutions are already possible in open
Dicke models~\cite{Ritsch2013,Mivehvar2021}, and cavity-assisted Raman
transitions are established in effective two-level
settings~\cite{Dimer2007,Baden2014}.
In the present model the tensor structure determines which Zeeman states
constitute the ordered branches, with $\avg{\hat J_z}$ for circular pumping
and $\avg{\hat J_z^2}$ for linear pumping, and it supplies channel-fixed
quadratic light shifts and $m$-dependent Raman vertices that cannot be
represented by a single pseudospin transition.
The leaked field probes only the Raman-bright relaxation channel and
therefore cannot reconstruct the dark spin motion.
Together with the two spin moments, it identifies the ordered branches.

Several approximations delimit the conclusions.
First, the atoms are restricted to a common spatial orbital, so recoil-driven
density ordering and spatial spin textures are excluded.
Second, the product-state closure neglects finite-$N$ correlations and
$1/N$ fluctuations; the cat terminology here refers to the structure of the
single-spin $J=6$ wave function, not to an $N$-atom entangled cat.
Third, magnetic dipole interactions, spin-dependent trapping, and atomic
dissipation are omitted.
Consequently, cavity loss damps only the Raman-bright sector and cannot make
the linear-pump state a global attractor.
Finally, the quoted thresholds are specific to the chosen $J=6$ parameters.
What extends to other open-shell atoms is the
scalar--vector--tensor operator structure and the channel dependence of
$u_t/u_v$, not the numerical critical couplings.

\begin{acknowledgments}
This work is supported by the National Key R\&D Program of China
(Grant No. 2022YFA1405302), the National Natural Science Foundation of
China (Grants No. 12675031 and No. 12174358), and NSAF
(Grant No. U2330401).
\end{acknowledgments}

\appendix

\section{Wigner--Eckart ratios}
\label{app:we}

The interaction in Eq.~\eqref{eq:effective-H} is the ground-manifold matrix
element of the two-photon operator obtained by adiabatically eliminating a
single odd-parity manifold $J'$.
Rotational covariance, equivalently the Wigner--Eckart theorem, restricts
that operator to irreducible ranks $0$, $1$, and $2$, which appear as the
scalar, vector, and tensor blocks.
The reduced matrix elements depend only on $(J,J')$, so the ratios of the
operator coefficients $u_s$, $u_v$, and $u_t$ are channel-fixed.
Matching the rank-two and rank-one blocks onto Eq.~\eqref{eq:effective-H}
gives
\begin{equation}
\frac{u_t}{u_v}=
\begin{cases}
-2/(2J+3),&J'=J+1,\\
-2,&J'=J,\\
+2/(2J-1),&J'=J-1.
\end{cases}
\label{eq:app-ratios}
\end{equation}
The channels $J'=J\pm1$ are those used in the numerics.
The middle line is the remaining electric-dipole possibility, unused below.
For $^{168}\mathrm{Er}$ with $J=6\rightarrow7$, Eq.~\eqref{eq:app-ratios}
is $u_t/u_v=-2/15$, or $u_s:u_v:u_t=70:15:-2$ after restoring the scalar
piece.
Any open-shell species on the same $(J,J')$ shares these ratios; we do not
quote a separate polarizability symbol because $\alpha$ in this paper is
the cavity amplitude.

\section{Channel reversal and blue detuning}
\label{app:channels}

The Raman matrix elements of the reversed channels already appear in
Figs.~\ref{fig:c-mechanism}(a) and~\ref{fig:l-mechanism}(a).
This appendix classifies the four combinations of pump detuning and
electric-dipole channel for each polarization.
Blue detuning reverses the signs of $V_0$ and $NU_0$ while leaving the
Raman vertices unchanged.
The empty-cavity orientation tracks $\operatorname{sign}(V_0 u_t)$;
boundedness of the product-state free energy tracks
$\operatorname{sign}(NU_0 u_t)$.

Table~\ref{tab:c-four} collects the circular classes.
The working design of the main text is the upper-left cell,
shown in Fig.~\ref{fig:c-phase}: the empty cavity occupies
$\ket{m=-J}$, the Raman vertex of $\hat K_+^{(c)}$ is dark there, and a
first-order jump transfers $\avg{\hat J_z}$ to positive $m$.
The opposite diagonal cell is the same mechanism with both signs reversed.
On blue-detuned $J\rightarrow J-1$, $u_t/u_v=+2/11$, the empty state is
$\ket{m=+J}$, and a finite cavity field jumps $\avg{\hat J_z}$ to negative
$m$.
The first-order window survives: $\eta_{\mathrm{well}}$, $\eta_b$, and
$\eta_c$ remain split, and the free energy is bounded for every
$\delta_c>0$.

The off-diagonal cells are a different class.
There $V_0 u_t>0$ places the empty-cavity minimum inside the manifold,
at $m^\ast=-1/(2u_t)$, where $\hat K_+^{(c)}$ is already bright.
The empty cavity then has no finite local-stability window.
The same cells have $NU_0 u_t>0$, so the product-state free energy is
unbounded unless $\delta_c$ exceeds a scale set by
$(NU_0)u_t J^2$, the circular analogue of Eq.~\eqref{eq:l-bounded}.
Above that line the empty packet remains interior; below it the
occupation runs to the nearby pole.

\begin{table}[t]
\caption{Circular-pump classes under channel reversal and blue detuning.
The working design of the main text is the upper-left cell.}
\label{tab:c-four}
\centering
\small
\setlength{\tabcolsep}{3.5pt}
\begin{tabular}{@{}l p{0.36\columnwidth} p{0.36\columnwidth}@{}}
\hline
 & \raggedright $u_t<0$ ($J\to J+1$) &
 \raggedright $u_t>0$ ($J\to J-1$) \tabularnewline
\hline
red ($V_0>0$) &
\raggedright pole $m=-J$, dark vertex,\newline
first order, bounded &
\raggedright interior minimum, bright vertex,\newline
boundedness line \tabularnewline
blue ($V_0<0$) &
\raggedright interior minimum, bright vertex,\newline
boundedness line &
\raggedright pole $m=+J$, first order,\newline
bounded \tabularnewline
\hline
\end{tabular}
\end{table}

Table~\ref{tab:l-four} is the corresponding linear classification.
The working design is the upper-left cell, shown in
Fig.~\ref{fig:l-phase}: an even-$m$ $\pm x$ cat, a soft real quadrature,
and the boundedness line of Eq.~\eqref{eq:l-bounded}.
Reversing the channel at red detuning is not a sign change of
$\avg{\hat J_z}$.
For $J\rightarrow J+1$, $u_t/u_v=-2/15$, the empty-cavity easy axis moves
from $\pm x$ to $\pm y$.
The soft Hessian direction becomes $\operatorname{Im}\alpha$, and the
threshold remains continuous and finite.
Past that point the condensate is a $y$-polarized coherent state with
$\avg{\hat J_y}\simeq\pm J$, and the free energy remains bounded.
The empty state of that class is an even-$m$ $\pm y$ cat.

Blue detuning on a given linear channel produces the other red-detuned
class, because the empty-cavity axis tracks $\operatorname{sign}(V_0 u_t)$
while boundedness tracks $\operatorname{sign}(NU_0 u_t)$.
The Raman nodes still follow $u_t$, so the blue thresholds lie close to,
but are not identical to, those of the opposite red cell.
On blue $J\rightarrow J-1$ the empty state is an even-$m$ $\pm y$ cat and
the condensate is $y$ polarized; on blue $J\rightarrow J+1$ the empty state
is a $\pm x$ cat and the condensate is cat-like with weight at large
$\abs m$.
Those two blue cells are the swaps of the red row.

\begin{table}[t]
\caption{Linear-pump classes under channel reversal and blue detuning.
The working design of the main text is the upper-left cell
(Fig.~\ref{fig:l-phase}).
Blue detuning swaps the two red classes; the blue thresholds lie close
to those of the opposite red cell.}
\label{tab:l-four}
\centering
\small
\setlength{\tabcolsep}{3.5pt}
\begin{tabular}{@{}l p{0.36\columnwidth} p{0.36\columnwidth}@{}}
\hline
 & \raggedright $u_t>0$ ($J\to J-1$) &
 \raggedright $u_t<0$ ($J\to J+1$) \tabularnewline
\hline
red ($V_0>0$) &
\raggedright $\pm x$ cat, soft $\operatorname{Re}\alpha$,\newline
Eq.~\eqref{eq:l-bounded} &
\raggedright $\pm y$ cat, soft $\operatorname{Im}\alpha$,\newline
bounded \tabularnewline
blue ($V_0<0$) &
\raggedright $\pm y$ cat, soft $\operatorname{Im}\alpha$,\newline
bounded &
\raggedright $\pm x$ cat, soft $\operatorname{Re}\alpha$,\newline
boundedness line \tabularnewline
\hline
\end{tabular}
\end{table}


\begin{thebibliography}{99}

\bibitem{Dicke1954}
R. H. Dicke,
Phys. Rev. \textbf{93}, 99 (1954).

\bibitem{HeppLieb1973}
K. Hepp and E. H. Lieb,
Ann. Phys. (N.Y.) \textbf{76}, 360 (1973).

\bibitem{Emary2003}
C. Emary and T. Brandes,
Phys. Rev. E \textbf{67}, 066203 (2003).

\bibitem{Baumann2010}
K. Baumann, C. Guerlin, F. Brennecke, and T. Esslinger,
Nature \textbf{464}, 1301 (2010).

\bibitem{Baumann2011}
K. Baumann, R. Mottl, F. Brennecke, and T. Esslinger,
Phys. Rev. Lett. \textbf{107}, 140402 (2011).

\bibitem{Mottl2012}
R. Mottl, F. Brennecke, K. Baumann, R. Landig, T. Donner, and T. Esslinger,
Science \textbf{336}, 1570 (2012).

\bibitem{Brennecke2013}
F. Brennecke, R. Mottl, K. Baumann, R. Landig, T. Donner, and T. Esslinger,
Proc. Natl. Acad. Sci. U.S.A. \textbf{110}, 11763 (2013).

\bibitem{Inouye1999}
S. Inouye, A. P. Chikkatur, D. M. Stamper-Kurn, J. Stenger, D. E. Pritchard, and W. Ketterle,
Science \textbf{285}, 571 (1999).

\bibitem{Moore1999}
M. G. Moore and P. Meystre,
Phys. Rev. Lett. \textbf{83}, 5202 (1999).

\bibitem{Keeling2014}
J. Keeling, M. J. Bhaseen, and B. D. Simons,
Phys. Rev. Lett. \textbf{112}, 143002 (2014).

\bibitem{Piazza2014b}
F. Piazza and P. Strack,
Phys. Rev. Lett. \textbf{112}, 143003 (2014).

\bibitem{Chen2014b}
Y. Chen, Z. Yu, and H. Zhai,
Phys. Rev. Lett. \textbf{112}, 143004 (2014).

\bibitem{Piazza2014a}
F. Piazza and P. Strack,
Phys. Rev. A \textbf{90}, 043823 (2014).

\bibitem{Chen2015}
Y. Chen, Z. Yu, and H. Zhai,
Phys. Rev. A \textbf{90}, 053609 (2015).

\bibitem{Zhang2021}
X. Zhang, Y. Chen, Z. Wu, J. Wang, J. Fan, S. Deng, and H. Wu,
Science \textbf{373}, 1359 (2021).

\bibitem{Deb2021}
A. B. Deb and N. Kj{\ae}rgaard,
Science \textbf{374}, 972 (2021).

\bibitem{Margalit2021}
Y. Margalit, Y.-K. Lu, F. {\c C}. Top, and W. Ketterle,
Science \textbf{374}, 976 (2021).

\bibitem{Sanner2021}
C. Sanner, L. Sonderhouse, R. B. Hutson, L. Yan, W. R. Milner, and J. Ye,
Science \textbf{374}, 979 (2021).

\bibitem{Klinder2015b}
J. Klinder, H. Ke{\ss}ler, M. R. Bakhtiari, M. Thorwart, and A. Hemmerich,
Phys. Rev. Lett. \textbf{115}, 230403 (2015).

\bibitem{Landig2016}
R. Landig, L. Hruby, N. Dogra, M. Landini, R. Mottl, T. Donner, and T. Esslinger,
Nature \textbf{532}, 476 (2016).

\bibitem{Bhaseen2009}
M. J. Bhaseen, M. Hohenadler, A. O. Silver, and B. D. Simons,
Phys. Rev. Lett. \textbf{102}, 135301 (2009).

\bibitem{Silver2010}
A. O. Silver, M. Hohenadler, M. J. Bhaseen, and B. D. Simons,
Phys. Rev. A \textbf{81}, 023617 (2010).

\bibitem{LeBoite2013}
A. Le Boite, G. Orso, and C. Ciuti,
Phys. Rev. Lett. \textbf{110}, 233601 (2013).

\bibitem{Li2013}
Y. Li, L. He, and W. Hofstetter,
Phys. Rev. A \textbf{87}, 051604(R) (2013).

\bibitem{Bakhtiari2015}
M. R. Bakhtiari, A. Hemmerich, H. Ritsch, and M. Thorwart,
Phys. Rev. Lett. \textbf{114}, 123601 (2015).

\bibitem{Hruby2018}
L. Hruby, N. Dogra, M. Landini, T. Donner, and T. Esslinger,
Proc. Natl. Acad. Sci. U.S.A. \textbf{115}, 3279 (2018).

\bibitem{Li2021}
X. Li, D. Dreon, P. Zupancic, A. Baumgartner, A. Morales, W. Zheng, N. Cooper, T. Donner, and T. Esslinger,
Phys. Rev. Research \textbf{3}, L012024 (2021).

\bibitem{Chen2016}
Y. Chen, Z. Yu, and H. Zhai,
Phys. Rev. A \textbf{93}, 041601 (2016).

\bibitem{Dogra2016}
N. Dogra, F. Brennecke, S. D. Huber, and T. Donner,
Phys. Rev. A \textbf{94}, 023632 (2016).

\bibitem{Sundar2016}
B. Sundar and E. J. Mueller,
Phys. Rev. A \textbf{94}, 033631 (2016).

\bibitem{Chen2022}
Y. Chen,
Commun. Theor. Phys. \textbf{74}, 125701 (2022).

\bibitem{Kollar2017}
A. J. Koll{\'a}r, A. T. Papageorge, V. D. Vaidya, Y. Guo, J. Keeling, and B. L. Lev,
Nat. Commun. \textbf{8}, 14386 (2017).

\bibitem{Leonard2017}
J. L{\'e}onard, A. Morales, P. Zupancic, T. Esslinger, and T. Donner,
Nature \textbf{543}, 87 (2017).

\bibitem{Ritsch2013}
H. Ritsch, P. Domokos, F. Brennecke, and T. Esslinger,
Rev. Mod. Phys. \textbf{85}, 553 (2013).

\bibitem{Mivehvar2021}
F. Mivehvar, F. Piazza, T. Donner, and H. Ritsch,
Adv. Phys. \textbf{70}, 1 (2021).

\bibitem{Klinder2015}
J. Klinder, H. Ke{\ss}ler, M. Wolke, L. Mathey, and A. Hemmerich,
Proc. Natl. Acad. Sci. U.S.A. \textbf{112}, 3290 (2015).

\bibitem{Keeling2010}
J. Keeling, M. J. Bhaseen, and B. D. Simons,
Phys. Rev. Lett. \textbf{105}, 043001 (2010).

\bibitem{Bhaseen2012}
M. J. Bhaseen, J. Mayoh, B. D. Simons, and J. Keeling,
Phys. Rev. A \textbf{85}, 013817 (2012).

\bibitem{Song2025}
E. Y. Song \textit{et al.},
Sci. Adv. \textbf{11}, eadu5799 (2025).

\bibitem{Gong2018}
Z. Gong, R. Hamazaki, and M. Ueda,
Phys. Rev. Lett. \textbf{120}, 040404 (2018).

\bibitem{Kessler2021}
H. Ke{\ss}ler, P. Kongkhambut, C. Georges, L. Mathey, J. G. Cosme, and A. Hemmerich,
Phys. Rev. Lett. \textbf{127}, 043602 (2021).

\bibitem{Kongkhambut2022}
P. Kongkhambut, J. Skulte, L. Mathey, J. G. Cosme, A. Hemmerich, and H. Ke{\ss}ler,
Science \textbf{377}, 670 (2022).

\bibitem{Lu2011}
M. Lu, N. Burdick, S. Youn, and B. Lev,
Phys. Rev. Lett. \textbf{107}, 190401 (2011).

\bibitem{Aikawa2012}
K. Aikawa \textit{et al.},
Phys. Rev. Lett. \textbf{108}, 210401 (2012).

\bibitem{Chomaz2023}
L. Chomaz \textit{et al.},
Rep. Prog. Phys. \textbf{86}, 026401 (2023).

\bibitem{Seo2024}
B. Seo, M. Huang, Z. Chen, M. K. Parit, Y. He, P. Chen, and G.-B. Jo,
Phys. Rev. Research \textbf{6}, L042028 (2024).

\bibitem{Eustice2020}
S. Eustice, K. Cassella, and D. Stamper-Kurn,
Phys. Rev. A \textbf{102}, 053327 (2020).

\bibitem{Schrott2024}
J. Schrott, D. Novoa, S. Eustice, and D. M. Stamper-Kurn,
Rev. Sci. Instrum. \textbf{95}, 113201 (2024).

\bibitem{Eustice2025}
S. Eustice \textit{et al.},
Phys. Rev. Research \textbf{7}, 023025 (2025).

\bibitem{Geremia2006}
J. M. Geremia, J. K. Stockton, and H. Mabuchi,
Phys. Rev. A \textbf{73}, 042112 (2006).

\bibitem{DeutschJessen2010}
I. H. Deutsch and P. S. Jessen,
Opt. Commun. \textbf{283}, 681 (2010).

\bibitem{LeKien2013}
F. Le Kien, P. Schneeweiss, and A. Rauschenbeutel,
Eur. Phys. J. D \textbf{67}, 92 (2013).

\bibitem{Lepers2014}
M. Lepers, J.-F. Wyart, and O. Dulieu,
Phys. Rev. A \textbf{89}, 022505 (2014).

\bibitem{Becher2018}
J. Becher \textit{et al.},
Phys. Rev. A \textbf{97}, 012509 (2018).

\bibitem{Dimer2007}
F. Dimer, B. Estienne, A. S. Parkins, and H. J. Carmichael,
Phys. Rev. A \textbf{75}, 013804 (2007).

\bibitem{Baden2014}
M. P. Baden, K. J. Arnold, A. L. Grimsmo, S. Parkins, and M. D. Barrett,
Phys. Rev. Lett. \textbf{113}, 020408 (2014).

\end{thebibliography}
\end{document}